# Feasibility Study of Internal Conversion Electron Spectroscopy of $^{229m}$Th

Benedict Seiferle[1a], Lars von der Wense[1], and Peter G. Thirolf[1]

Ludwig-Maximilians-Universität München, Am Coulombwall 1, Garching, Germany

February 1$^{\text{st}}$, 2017

**Abstract.** With an expected energy of 7.8(5) eV, the isomeric first excited state in $^{229}$Th exhibits the lowest excitation energy of all known nuclei. Until today, a value for the excitation energy has been inferred only by indirect measurements. In this paper, we propose to use the internal conversion decay channel as a probe for the ground-state transition energy. MatLab-based Monte Carlo simulations have been performed to obtain an estimate of the expected statistics and to test the feasibility of the experiment. From the simulations we conclude, that with the presented methods an energy determination with a precision of better than 0.1 eV is possible.

## 1 Introduction

The isomeric first excited state of $^{229}$Th, called $^{229m}$Th, is subject to current vivid research. Among all known nuclear excited states it is the only one that could allow for a direct optical laser excitation, due to its extraordinary low excitation energy of only 7.8(5) eV, corresponding to about 160 nm [1,2]. This has led to a multitude of proposals for possible applications, including a nuclear optical clock [3,4], that could provide a complementary technology to today's existing optical atomic clocks, potentially even outperforming the present frequency standards due to the superior resilience of a nuclear clock against external perturbations. It took 40 years until the first direct identification of the ground-state decay of $^{229m}$Th via the observation of its internal conversion decay branch [5]. However, despite large experimental efforts conducted worldwide [6–13], the uncertainty in the excitation energy value is still too large to allow for a direct laser excitation. By now, energy values have only been acquired with indirect measurements, investigating nuclear excited states at higher energies and $\gamma$ rays emitted in their decays to the ground- and isomeric state [1,2,14–16].

There are three decay channels of $^{229m}$Th to its ground state discussed in literature [17,18]: (i) internal conversion (IC), which proceeds via the emission of an electron with an energy of $E_e = E_I - E_B$, where $E_I$ is the isomeric energy and $E_B$ is the binding energy of the electron. (ii) $\gamma$ decay, where the emitted photon carries the energy of the isomer, and (iii) bound internal conversion, which proceeds via the excitation of a bound electronic shell state.

While a variety of proposals and experimental attempts can be found in literature, aiming at the direct measurement of a VUV photon emitted during the ground-state decay of $^{229m}$Th [6,7,19], this paper investigates the possibilities of an energy determination via the electron that is emitted during the already experimentally observed internal conversion decay [5,20].

---

[a] *Present address:* benedict.seiferle@physik.uni-muenchen.de

Internal conversion electron spectroscopy of $^{229m}$Th has several advantages compared to the photonic approach: Due to the large conversion coefficient, the decay via internal conversion is about $10^9$ times faster than the photonic decay. Therefore it is possible to trigger the IC decay by neutralizing a $^{229m}$Th ion: IC decay is only possible, if the binding energy of an electron in the surrounding of the nucleus is below the isomeric energy. Therefore IC is suppressed in $^{229m}$Th ions, but not in the neutral thorium atom [21].

## 2 Simulated Setup

In our experimental setup, $^{229(m)}$Th ions are produced as $\alpha$ recoil ions from a thin extended $^{233}$U source. A decay branch of 2% ends up in the isomeric first excited state of $^{229}$Th. $^{229(m)}$Th ions are stopped in a buffer-gas stopping cell filled with ultra-pure helium, so that after thermalization of the $\alpha$-recoil ions in the buffer gas and their transport via RF- and DC fields to an extraction nozzle $^{229(m)}$Th ions can be extracted into a segmented radio frequency quadrupole (RFQ) ion guide and phase-space cooler structure. The segmented structure of the RFQ allows to form ion bunches. A subsequent quadrupole mass separator removes accompanying $\alpha$-decay daughter products. A detailed study of the experimental setup can be found in [22]. A width of the ion bunches of 10 $\mu$s (FWHM$_{\text{TOF}}$), with $\approx$200 $^{229(m)}$Th$^{3+/2+}$ ions per bunch was achieved at a rate of 10 Hz [20]. The ion bunches contain $^{229m}$Th, which has already been detected with this setup [5].

The general idea of performing internal conversion electron spectroscopy of $^{229m}$Th is to guide the extracted, mass-separated and bunched ions towards an electron spectrometer, where they will be neutralized, thereby triggering the IC decay and to measure the subsequently emitted electrons.

In this work, an approach is studied, where $^{229m}$Th ions are collected directly on a metallic catcher for neutralization. As there are only about 4 ions in the isomeric state extracted per bunch, it is advantageous to use a spectrometer with a high efficiency. Therefore, a magnetic bottle-type spectrometer [23], providing an acceptance angle of nearly $4\pi$, is envisaged. In such a spectrometer, electrons are collected and collimated by a magnetic gradient field. In general, the electron energy is then either inferred by a time-of-flight method or by retarding fields. In case of $^{229m}$Th, the time-of-flight method cannot be used, since the lifetime of the internal conversion decay is roughly 10 $\mu$s and thus long compared to the short flight time of electrons ($v \approx 6 \times 10^5$ m/s for 1 eV electrons). By applying retarding fields, an integrated spectrum is generated, where all electrons are counted, whose energy is sufficient to pass the retarding fields that are applied with high-transmission grids. In the following, not the specific features of the magnetic bottle spectrometer are investigated, but rather measurement principles are discussed to check the feasibility of high-precision internal conversion electron spectroscopy of the $^{229}$Th isomer.

### 2.1 Solid Sample & Surface Effects

A simple way to neutralize the $^{229m}$Th ions is to collect them on a metallic catcher. If the implantation depth is not too deep[1], electrons emitted during the decay should be able to leave the sample and be measured by the spectrometer. A possible source of background for experiments with slow ions is electron emission via Auger processes (see [25]). When ion bunches are used, these electrons can be distinguished from electrons emitted by the isomeric decay: Due to its lifetime of $\approx$ 10 $\mu$s, the isomeric decay can be temporally separated from signals potentially generated by ionic impact [20].

It can be expected that the energy distribution of the electrons reflects the electronic structure of the catcher's surface. The processes are similar to metastable atom electron spectroscopy

---
[1] The stopping range for Thorium ions with $E_{\text{kin}} = 100$ eV (500 eV) in gold is 5 Å (8 Å) (values are taken from SRIM simulations [24]).

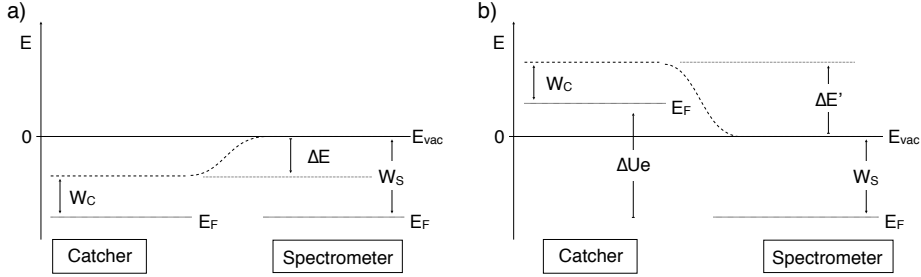

Fig. 1: Visualization of the work function of the catcher and spectrometer (see also [29]). Two situations are shown: a) Contact potential difference $\Delta E$ generated between the catcher and the spectrometer. b) Influence of an applied offset voltage $\Delta U$ on the contact potential difference.

(MAES) [26,27], that is used to study the electronic structure of surfaces.

### 2.1.1 Surface Influence

In this section, the influence of the catcher material (or sample) is investigated and a possible measurement scheme is shown.

The problem of measuring the isomer's energy is similar, but not identical to ultraviolet photo electron spectroscopy, where a surface is irradiated with UV photons of known energy [28]. The electronic structure is then inferred from the energy of electrons emitted during the photoelectric effect. Opposed to that, when measuring the isomeric energy, it is the objective to infer the energy of the "light source" (*i.e.* isomer) from the energy distribution of the electrons. In the following, a short review of the terms and measurement schemes deployed in photo electron spectroscopy is given (see also [29]):

The work function $W$ of a metallic material is defined as the potential energy difference between the local vacuum level ($E_{\text{vac}}$) and the Fermi level ($E_F$):

$$W = E_{\text{vac}} - E_F. \quad (1)$$

When two materials (for example the spectrometer and the catcher surface with work functions $W_S$ and $W_C$) are in electrical contact, their Fermi levels align. If their work functions differ, a potential difference between the local vacuum levels is generated. The contact potential difference amounts to

$$\Delta E = W_C - W_S. \quad (2)$$

In our situation, a contact potential difference may be generated between the catcher and the spectrometer, which is visualized in Fig. 1a. Therefore, if the work function of the spectrometer exceeds the catcher's work function, an offset voltage $\Delta U$ needs to be applied to the sample in order to give the electrons enough energy to overcome the contact potential difference. Consequently, the contact potential difference is shifted by $\Delta U$ (see Fig. 1b):

$$\Delta E' = W_C - W_S + \Delta U e. \quad (3)$$

Photons of energy $h\nu$ may eject electrons from a metallic surface with a work function $W_C$, as long as $h\nu \geq W_C$. The energy of such a photo electron is described by a Fermi distribution with a maximum energy of

$$E_C^{\max} = h\nu - W_C. \quad (4)$$

Note, that in this definition $E_C^{\max}$ is given with respect to the local vacuum energy level of the catcher. Given the shifted contact potential difference with the spectrometer, $\Delta E'$, the maximum kinetic energy of the electrons measured with the spectrometer amounts to

$$E_S^{\max} = h\nu - W_C + \Delta E' \qquad (5)$$
$$= h\nu - W_C + W_C - W_S + \Delta Ue \qquad (6)$$
$$= h\nu - W_S + \Delta Ue. \qquad (7)$$

From the above equation it is obvious, that the energy of the electrons in the end does not depend on the value of the work function of the catcher, but only on the spectrometer work function and the applied offset voltage $\Delta U$. Treating the isomeric decay of $^{229m}$Th as a photon with energy $E_I = h\nu$ that is coupling to the electrons in the catcher surface, the energy of the isomer $E_I$ can be inferred by the following equation:

$$E_I = E_S^{\max} + (W_S - \Delta Ue), \qquad (8)$$

where the expression $(W_S - \Delta Ue)$ can be measured with a light source of known energy and using Eq. (7). Therefore, the only remaining surface influence of the sample on the maximum kinetic energy of an electron is the temperature dependent Fermi distribution of $E_e^{\max}$, but not the value of the sample work function.

## 3 Simulations

In order to get an estimate for the count rates and resulting integrated spectra that can be measured for $^{229m}$Th IC electrons emitted from a solid sample, Monte Carlo (MC) simulations were performed with a custom MC code.

### 3.1 Generation Of Density Distribution

The code allows to simulate a predefined kinetic energy distribution $D(E)$ of the electrons. To perform a Monte Carlo simulation, an arbitrary uniform random number $p_i \in [0, 1]$ is mapped to a kinetic energy $E_i$. The abundance of particles with energies $E$ should then finally reflect the kinetic energy distribution $D(E)$. We achieve this by performing the following operations:
At first, $E \in [0, E_{\max}]$ is discretized into $N + 1$ parts, equally spaced with $\Delta E = E_{\max}/N$. For simplicity we define

$$E_i = (0.5 + i) \cdot \Delta E, \text{ and} \qquad (9)$$
$$D_i = D(E_i). \qquad (10)$$

We take $S_k = \sum_{i=0}^{k} D_i$, with $0 < k \leq N$, calculate the sum $I = \sum_{k=0}^{N} S_k$ and define the normalized $\hat{S}_i = (1/I) \cdot S_i$. In this way, $\hat{S}_i$ is a number between 0 and 1. If now a random number $p \in [0, 1]$ lies between $\hat{S}_i$ and $\hat{S}_{i+1}$, then the output energy $E$ should be an arbitrary value between $(E_i - 0.5\Delta E)$ and $(E_i+0.5\Delta E)$: $E = E_i+(0.5-r)\cdot\Delta E$, where $r$ is a uniform random number $\in [0, 1]^2$. In this way it is possible to map a random number between 0 and 1 to an energy value. The procedure is visualized with an example in Fig. 2, where we used a Gaussian energy distribution $(D(E) = a \cdot \exp(-(E-\mu)^2/(2\sigma^2)))$ with $\mu = 2$ and $\sigma = 0.5$: $D(E) = \frac{2}{\sqrt{2\pi}} \exp\left[-2(E-2)^2\right]$. To define the values of $\hat{S}_i$, the following parameters were used: $N = 20$ and $E_{\max} = 4$. In the following simulations, the values $N = 300$ and $E_{\max} = 4$ eV were used and a superposition of a Gaussian distribution with a Fermi distribution was used as input functions.

### 3.2 Description of the simulation process

In the following section, the simulation process is discussed. The number of isomers per bunch is calculated by taking the 2% branching ratio to the isomeric state from the $^{233}$U $\alpha$ decay for the 200 ions contained in one bunch. In this way, we are left with 4 ions in the isomeric state per bunch. We further assume that only 20% of the ions are collected in the center of the spectrometer and contribute to a spectrum. A collimation efficiency of the magnetic field of 80%, a combined grid transmission of 50% (3 grids with 80% geometrical transmission each) and a detection efficiency of 30% was used. Since a catcher surface is used, only 50% of the IC electrons that are potentially emitted in one hemisphere can be collected. In this way we are left with a total combined detection efficiency of the spectrometer of $\epsilon = 6\%$. General input values that were used for the simulation, such as the detection properties, count rates, resolution and temperature are listed in Table 1. Background was simulated by calculating the signal-to-background ratio and simulating the dark counts accordingly.

---

[2] Note that the values for $S_i$ can also be generated analytically, as long as $D(E)$ stays doubly integrable (which is not the case for a Gaussian distribution). Then $S(E) = \int\limits_0^E dE' D(E')$ and $I = \int\limits_0^\infty dE' S(E')$.

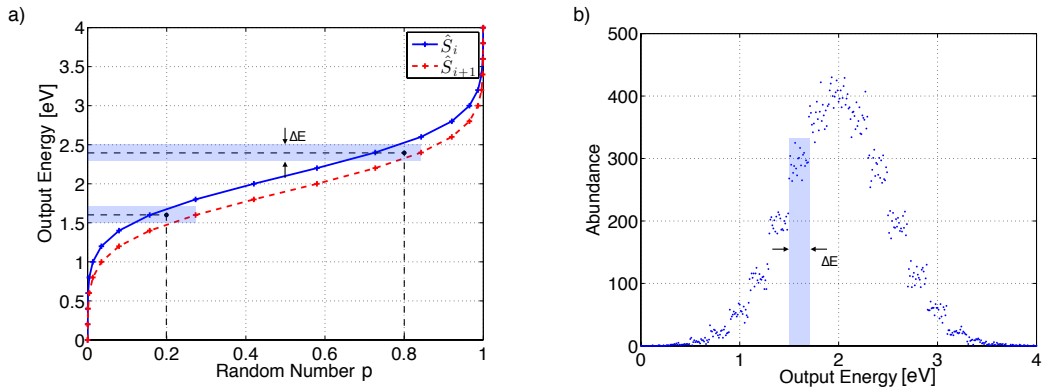

Fig. 2: Visualization of the energy distribution generation. a) shows the curves $\hat{S}_i$ and $\hat{S}_{i+1}$ as defined in the main text. Two possible values of the random numbers $p$ are also indicated by two vertical black dashed lines: In this case $p = 0.2$ would lead to an output energy between 1.5 eV and 1.7 eV. $p = 0.8$ results in an output energy between 2.3 eV and 2.5 eV. b) shows the resulting distribution for 50,000 randomly generated values of $p$ between 0 and 1. Since $N$ was chosen to be relatively small, $\Delta E = 0.2$ eV is large and the discretized values of $E$ can be directly seen in this plot.

As already mentioned in Sect. 2.1.1, the maximum energy of the electrons (with respect to the spectrometer) does not depend on the work function of the sample, but rather on the work function of the spectrometer and the sample offset voltage $(W_S - \Delta Ue)$, that needs to obtained from a calibration measurement with a light source of known energy. Nevertheless, the work function of the sample does play a role, since $h\nu = E_I \geq W_C$ must always be satisfied and the electron energies (with respect to the vacuum level of the sample) are distributed between 0 and $(E_I - W_C)$ eV. As a typical work function of metals, $W_C$ was set to 5 eV. In the simulations $(W_S - \Delta Ue)$ was set to be equal to 5 eV (this can result no contact potential difference and 0 V offset voltage). In this way, the electron energies are distributed over a range between 0 and $E_I - 5$ eV. For the isomer energy $E_I$ two values were simulated: 7.8 eV and 7.9 eV. The two values with a difference of 0.1 eV were chosen in order to check the resolving power of this approach. Since the photoelectrons reflect the surface's electronic structure, one cannot assume a "bare" Fermi distribution in the simulation. This is taken into account by adding a Gaussian energy distribution to the low-energy part of the electron spectrum, so that only $\approx$ 10% of the electrons have a higher energy than 2 eV. The simulated energy distribution and resulting spectra are shown in Fig. 3.

## 4 Analysis Of Simulated Spectra

Spectra measured with retarding field analyzers are typically differentiated to gain information on the electronic structure of the surfaces. Since we are only interested in the maximum energy of the electron (*i.e.* the Fermi edge) and a differentiation may lead to large relative errors, we directly fit the indefinite integral of the Fermi function to the high energy part of the integrated spectrum.

| spectrometer efficiency | % | simulation input | | | | |
|---|---|---|---|---|---|---|
| collimation | 80 | MCP dark count rate [1/s] | 35 | collection efficiency [%] | 20 |
| grid transmission | 50 | bunches per second [1/s] | 10 | resolution (FWHM) [eV] | 0.1 |
| MCP detection | 30 | ions per bunch | 200 | T [K] | 300 |
| hemisphere | 50 | read out time per bunch [$\mu$s] | 200 | | |
| Total: | 6 | | | | |

Table 1: List of input values that were used for the simulations as discussed in the text and shown in the plots (Fig. 3, 4 and 5). The read-out time is the width of the time window in which counts of the isomer are expected and read out.

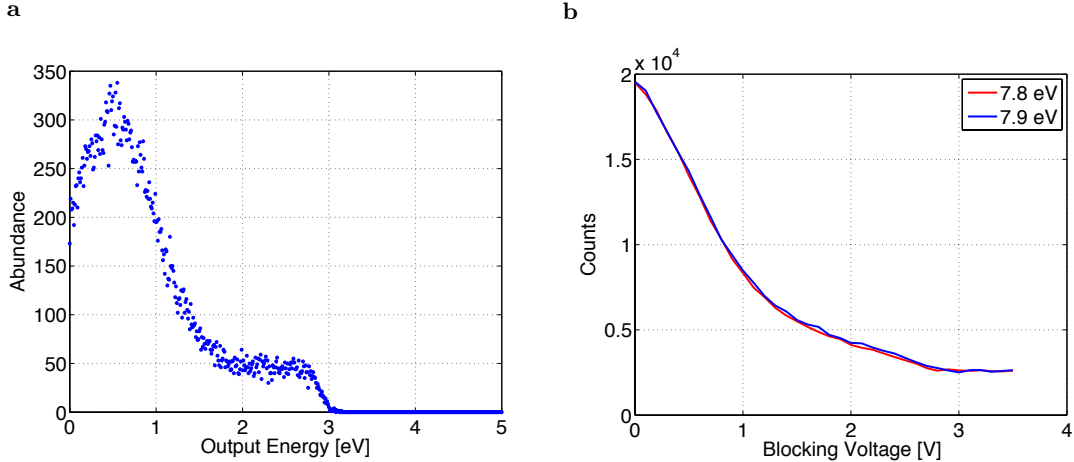

Fig. 3: Simulated spectra: a) The simulated energy distributions for a "bare" Fermi distribution with some arbitrary additional underlying electronic structure. b) Corresponding results from a simulated measurement with the retarding field spectrometer. A measurement time of 360 h and a blocking voltage increment of 0.1 V was chosen.

When a solid sample is used for the neutralization of the $^{229m}$Th ions, the subsequently emitted electrons reflect the electronic structure of the surface (see sect. 2.1.1). Especially the maximum energy edge must reflect the Fermi distribution:

$$f(E) = \frac{a}{e^{((E-E_0)/b)} + 1}, \quad (11)$$

with $b = k_B \cdot T$, $E_0$ as the maximum kinetic energy of the electrons and $a$ as a constant. Its antiderivative reads

$$F(E) = a \cdot \left( b \cdot \ln \left[ \frac{e^{((E-E_0)/b)} + 1}{e^{-E_0/b} + 1} \right] - E \right) + C, \quad (12)$$

where $C$ is a constant. Fig. 4 shows the simulation with voltage increments of 0.04 V over a range of 1.5 V and a measurement time of 180 h (leading to 38 data points, with 4.8 h measurement time per data point) and the corresponding fit plots. For 7.8 eV (7.9 eV) isomeric energy, a maximum kinetic energy of $E = 2.80 \pm 0.05$ eV

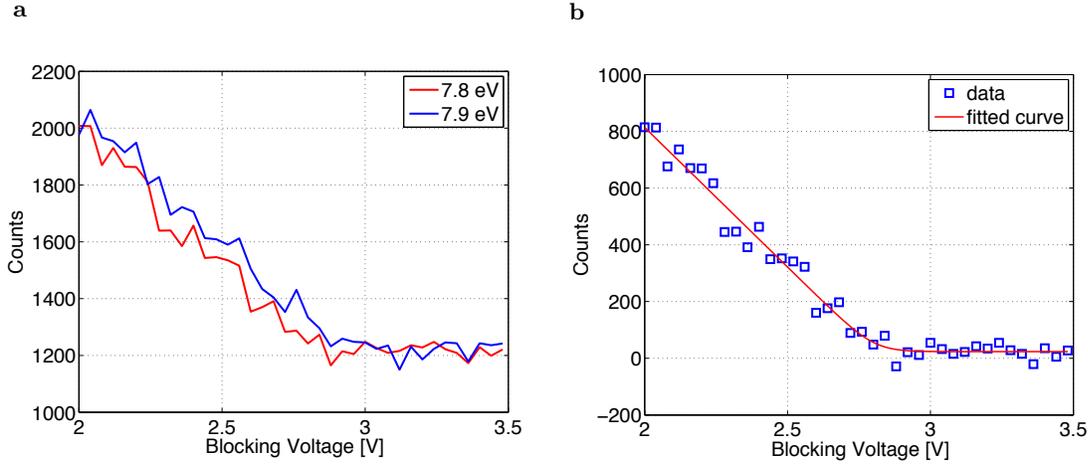

Fig. 4: a) Simulated spectra for two isomeric energies (7.8 eV and 7.9 eV) with a measurement time of 180 h, blocking voltage increment of 0.04 V in a range of 1.5 V (leading to 4.8 h measurement time per data point), $W_C = 5$ eV and $(W_S - \Delta Ue) = 5$ eV. b) One of the fitted curves (according to Eq. (12)). The fit values were $E = 2.80 \pm 0.05$ eV (for 7.8 eV) and $E = 2.91 \pm 0.05$ eV (for 7.9 eV). Note that the calculated background (1210 counts) was subtracted in the fit plot.

($E = 2.91 \pm 0.05$ eV) was obtained from the fit.

## 5 Energy resolution

The precision and accuracy of the fit method was probed, by performing 1000 simulations (each with a specific measurement time (90 h, 180 h, 270 h), 1.5 V blocking voltage range and blocking voltage increment of 0.04 V). The fit results of the maximum kinetic energy value were then subtracted from the simulated energy and the difference was filled in a histogram. The histograms and corresponding Gaussian fits[3] are plotted in Fig. 5. The fit results are shown in Table 2. Taking these results, it is obvious that the precision and accuracy both improve with longer measurement times. The width $\sigma$ follows a $\sqrt{N}$ law (starting below 50 meV for a measurement time of 90 h). Although it is much smaller than the width of the distribution, there is a shift towards lower energies, which is decreasing with better statistics.

---

[3] $f(x) = a \cdot \exp(-(x-\mu)^2/(2\sigma^2))$ was used for the fit function.

| meas. time [h] | $\sigma$ [meV] | $\mu$ [meV] |
|---|---|---|
| 90 | 33 | −8 |
| 180 | 22 | −7 |
| 270 | 18 | −3 |

Table 2: Fit results for the function $f(x) = a \cdot \exp(-(x-\mu)^2/(2\sigma^2))$ fitted to the curves shown in Fig. 5.

## 6 Conclusion and Outlook

We presented a way to measure the excitation energy of the isomeric first excited state in $^{229}$Th via internal conversion electrons. The approach uses a metallic catcher to neutralize $^{229m}$Th ions to open the IC decay channel. The analysis of simulated data results in uncertainties of below 0.1 eV in a reasonable measurement time of 180 h ($\hat{=}$ 7.5 d).

One needs to mention, that there is no influence of the sample material on the absolute achieved energy value, since only the maximum kinetic energies of the electrons are measured and no specific binding energies of electrons in the sample. Therefore, the cleanliness of the sample surface does not affect the energy measurements. Still different metallic materials can be probed to enhance confidence in the obtained energy value and investigate systematic shifts. Taking all together, we conclude that it is possible to measure the isomeric energy to better than 0.1 eV with the proposed method.

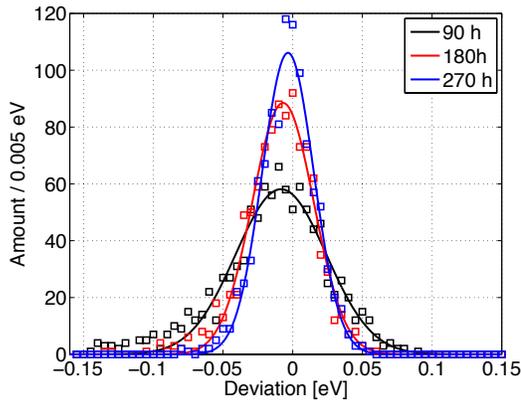

Fig. 5: Precision and accuracy of the fit method. The deviation from the simulated energy value for 1000 measurements, with different measurement times (90 h (black), 180 h (red), 270 h (blue)) is shown. The results for the corresponding Gaussian fits are shown in Table 2.

We acknowledge fruitful discussions with S. Stellmer, M. Laatiaoui, P. Feulner, G. Dedes and J. Crespo López-Urrutia. This work was supported by DFG grant (Th956/3-1), via the European Union's Horizon 2020 research and innovation programme under grant agreement No. 664732 "nuClock" and by the LMU department of Medical Physics via the Maier-Leibnitz Laboratory.